\documentclass[10pt,aps,prd,showpacs,groupedaddress,nofootinbib,notitlepage,preprintnumbers,twocolumn]{revtex4-2}

\usepackage{braket,bm}
\usepackage[pass]{geometry} 
\usepackage{color,graphicx}
\usepackage{bm}        
\usepackage{amsmath,amsfonts,amssymb}   
\usepackage{mathrsfs,mathtools, wasysym}
\DeclareMathAlphabet{\mathpzc}{OT1}{pzc}{m}{it}
\usepackage{hyperref}

\begin{document}

%%%%%%%%%%%%%%%%%%%%%%%%%%%%%%%%%%%%%%%%%%%%%%%%%%%%%%%%%%%%%%%%%%%%%%
\title{Alternative flow equation for the\\ functional renormalization group}

\author{Elizabeth Alexander}
\affiliation{School of Physics and Astronomy, University of Nottingham,\\ Nottingham NG7 2RD, United Kingdom}

\author{Peter Millington}
\email{p.millington@nottingham.ac.uk}
\affiliation{School of Physics and Astronomy, University of Nottingham,\\ Nottingham NG7 2RD, United Kingdom}

\author{Jordan Nursey}
\affiliation{School of Physics and Astronomy, University of Nottingham,\\ Nottingham NG7 2RD, United Kingdom}

\author{Paul M. Saffin}
\email{paul.saffin@nottingham.ac.uk}
\affiliation{School of Physics and Astronomy, University of Nottingham,\\ Nottingham NG7 2RD, United Kingdom}

\date{July 28, 2021}

%%%%%%%%%%%%%%%%%%%%%%%%%%%%%%%%%%%%%%%%%%%%%%%%%%%%%%%%%%%%%%%%%%%%%%
\begin{abstract}
We derive an alternative to the Wetterich-Morris-Ellwanger equation by means of the two-particle irreducible (2PI) effective action, exploiting the method of external sources due to Garbrecht and Millington. The latter allows the two-point source of the 2PI effective action to be associated consistently with the regulator of the renormalization group flow. We show that this procedure leads to a flow equation that differs from that obtained in the standard approach based on the average one-particle irreducible effective action.\bigskip

\noindent\footnotesize{This is an author-prepared post-print of \href{https://doi.org/10.1103/PhysRevD.100.101702}{Phys.\ Rev.\ D 100 (2019) 101702(R)}, published by the American Physical Society under the terms of the \href{https://creativecommons.org/licenses/by/4.0/}{CC BY 4.0} license (funded by SCOAP\textsuperscript{3}).  This version has been amended to incorporate corrections described in an erratum, submitted in respect of the published article.}
\end{abstract}


\maketitle

\section{Introduction}

The effective action~\cite{Jackiw:1974cv, Cornwall:1974vz} provides a powerful framework for describing the nonperturbative behavior of quantum mechanical systems, having been employed extensively in both the relativistic and nonrelativistic regimes. Once extended by the introduction of a regulator, which allows us to integrate in only a continuous subset of momentum modes above a given energy scale $k$, the so-called average one-particle irreducible (1PI) effective action~\cite{Wetterich:1989xg} yields a self-consistent equation, due to Wetterich~\cite{Wetterich:1992yh}, Morris~\cite{Morris:1993qb} and Ellwanger~\cite{Ellwanger:1993mw} (see also Ref.~\cite{Reuter:1996cp} by Reuter in the context of gravity), for the renormalization group (RG) flow of the effective action (for reviews, see~Refs.~\cite{Berges:2000ew, Pawlowski:2005xe, Gies:2006wv, Rosten:2010vm}). This flow equation has been used to study critical phenomena~\cite{Tetradis:1993ts, Berges:1995mw, Litim:2002cf} (cf.~Ref.~\cite{Pelissetto:2000ek}), to illustrate the emergence of the Maxwell construction in theories with spontaneous symmetry breaking~\cite{Tetradis:1992qt} (cf.~Ref.~\cite{Alexandre:1998ts}), and to derive the beta functions and identify the fixed points of various interacting quantum field theories, notably in the context of the on-going asymptotic safety program of quantum gravity~\cite{Shaposhnikov:2009pv, Eichhorn:2010tb, Dietz:2012ic, Falls:2014tra, Falls:2018ylp} (for reviews, see Refs.~\cite{Niedermaier:2006wt, Codello:2008vh, Reuter:2012id, Eichhorn:2018yfc}), as initiated by Weinberg~\cite{WeinbergAS}.

In this article, we derive an alternative flow equation from the two-particle irreducible (2PI) effective action by means of the method of external sources due to Garbrecht and Millington~\cite{Garbrecht:2015cla}. We show that the resulting flow equation differs to that derived from the average 1PI effective action, suggesting there exists an ambiguity in the ``correct'' choice of exact flow equation. The procedure presented here does not amount to a 2PI generalization of the average 1PI effective action, cf., e.g., Refs.~\cite{Pawlowski:2005xe,Carrington:2014lba}.

\section{2PI effective action}

The 2PI effective action~\cite{Cornwall:1974vz}
\begin{equation}
\Gamma^{\rm 2PI}[\phi,\Delta]=W[\mathcal{J},\mathcal{K}]+\mathcal{J}_x\phi_x+\frac{1}{2}\mathcal{K}_{xy}\left(\phi_x\phi_y+\hbar\Delta_{xy}\right)
\end{equation}
is the Legendre transform of the Schwinger functional
\begin{equation}
W[\mathcal{J},\mathcal{K}]=-\hbar\ln Z[\mathcal{J},\mathcal{K}]
\end{equation}
with respect to the sources $\mathcal{J}$ and $\mathcal{K}$, where
\begin{equation}
\label{eq:Z}
Z[\mathcal{J},\mathcal{K}]=\int\!\mathcal{D}\Phi\;\exp\left[-\tfrac{1}{\hbar}\left(S[\Phi]-\mathcal{J}_z\Phi_z-\tfrac{1}{2}\mathcal{K}_{zw}\Phi_z\Phi_w\right)\right]
\end{equation}
is the source-dependent Euclidean path integral for the theory with classical action $S[\Phi]$. We employ the DeWitt notation throughout, wherein repeated continuous indices are integrated over, i.e.~\smash{$\mathcal{J}_x\phi_x\equiv\int\!{\rm d}^4x\,\mathcal{J}(x)\phi(x)$}.

The sources $\mathcal{J}$ and $\mathcal{K}$ are functionals of the conjugate variables $\phi$ and $\Delta$, i.e.~$\mathcal{J}_x\equiv\mathcal{J}_x[\phi,\Delta]$ and $\mathcal{K}_{xy}\equiv\mathcal{K}_{xy}[\phi,\Delta]$, defined via the partial functional variations
\begin{subequations}
\begin{align}
\label{eq:JplKdef}
\frac{\delta \Gamma^{\rm 2PI}[\phi,\Delta]}{\delta \phi_x}&=\mathcal{J}_x+\mathcal{K}_{xy}\phi_y,\\
\frac{\delta \Gamma^{\rm 2PI}[\phi,\Delta]}{\delta \Delta_{xy}}&=\frac{\hbar}{2}\mathcal{K}_{xy}.
\end{align}
\end{subequations}
The conjugate variables are respectively the connected one- and two-point functions
\begin{subequations}
\label{eq:vardefs}
\begin{align}
\phi_x&=-\frac{\delta W[\mathcal{J},\mathcal{K}]}{\delta \mathcal{J}_x},\\
\hbar\Delta_{xy}&=-2\frac{\delta W[\mathcal{J},\mathcal{K}]}{\delta \mathcal{K}_{xy}}-\phi_x\phi_y,
\end{align}
\end{subequations}
which are, in corollary, functionals of the sources $\mathcal{J}$ and $\mathcal{K}$, i.e.~$\phi_x\equiv\phi_x[\mathcal{J},\mathcal{K}]$ and $\Delta_{xy}\equiv\Delta_{xy}[\mathcal{J},\mathcal{K}]$.

We can proceed perturbatively by performing a saddle-point evaluation of the path integral in Eq.~\eqref{eq:Z}. The saddle points $\{\varphi\}$ satisfy the stationarity condition
\begin{equation}
\label{eq:stationarity}
\frac{\delta S[\Phi]}{\delta \Phi_x}\bigg|_{\Phi=\varphi}-\mathcal{J}_x[\phi,\Delta]-\mathcal{K}_{xy}[\phi,\Delta]\varphi_y=0,
\end{equation}
indicating that $\varphi$ is itself a functional of $\phi$ and $\Delta$, and thereby also $\mathcal{J}$ and $\mathcal{K}$, i.e.~$\{\varphi\}\equiv\{\varphi\}[\phi,\Delta]$. The placement of the functional arguments reflects the fact that both the number and nature of the saddle points depend on the configuration $(\phi,\Delta)$, see Ref.~\cite{Millington:2019nkw}.

In the approach of Ref.~\cite{Garbrecht:2015cla}, and in the case of a single saddle point, the stationarity condition in Eq.~\eqref{eq:stationarity}, combined with the variation in Eq.~\eqref{eq:JplKdef}, can be used to constrain the linear combination $\mathcal{J}_x+\mathcal{K}_{xy}\phi_y$ of the sources. This, however, provides only one constraint, and we are free to choose the other, fixing, for instance, the form of the two-point source $\mathcal{K}_{xy}$. If we choose this constraint to be the Schwinger-Dyson equation then we recover the standard Cornwall-Jackiw-Tomboulis 2PI effective action~\cite{Cornwall:1974vz}, with the exception that the saddle-point configuration is driven towards the quantum-corrected trajectory of the system. The latter feature is particularly relevant in the case of false vacuum decay in theories with radiatively generated spontaneous symmetry breaking (see Refs.~\cite{Weinberg:1992ds,Garbrecht:2015cla,Garbrecht:2015yza}), for instance via the Coleman-Weinberg mechanism~\cite{Coleman:1973jx}. Alternatively, we can constrain the two-point source to be local, i.e.~taking \smash{$\mathcal{K}_{xy}=\mathcal{K}_x\delta^{4}(x-y)$}, giving the two-particle point-irreducible (2PPI) effective action of Verschelde and Coppens~\cite{Verschelde:1992bs}. If, in the case of global symmetries, we instead use the Ward identities to constrain the two-point source in perturbative truncations of the effective action, we obtain results in the spirit of the symmetry-improved 2PI effective action of Pilaftsis and Teresi~\cite{Pilaftsis:2013xna}.

In this article, we will choose the two-point source to be the regulator of the RG evolution and find that this procedure does not reproduce the well-known flow equation due to Wetterich~\cite{Wetterich:1992yh}, Morris~\cite{Morris:1993qb} and Ellwanger~\cite{Ellwanger:1993mw}.

\section{Exact flow equations}

\subsection{Average-1PI approach}

The standard derivation of the exact flow equation follows from the average 1PI effective action
\begin{equation}
\label{eq:ave1PI}
\Gamma^{\rm 1PI}_{\rm av}[\phi,\mathcal{R}^{(k)}]=W[\mathcal{J},\mathcal{R}^{(k)}]+\mathcal{J}_x\phi_x+\frac{1}{2}\phi_x\mathcal{R}^{(k)}_{xy}\phi_y,
\end{equation}
where $\mathcal{J}_x\equiv \mathcal{J}_x[\phi]$ and
\begin{equation}
\label{eq:phidef}
\phi_x=-\frac{\delta W[\mathcal{J},\mathcal{R}^{(k)}]}{\delta \mathcal{J}_x}.
\end{equation}
The regulator\footnote{Note that we use an unusual sign convention for the definition of the regulator in order to make a clearer comparison with our 2PI approach.} $\mathcal{R}^{(k)}_{xy}$ appears in the path integral $Z[\mathcal{J},\mathcal{R}^{(k)}]$, as defined in Eq.~\eqref{eq:Z}, leading to the scale-dependent Schwinger functional $W[\mathcal{J},\mathcal{R}^{(k)}]\equiv-\hbar \ln Z[\mathcal{J},\mathcal{R}^{(k)}]$, whose variation with respect to the scale $k$ yields the Polchinski equation~\cite{Polchinski:1983gv}. Notice that no extremization is taken with respect to the regulator.

If $\phi$ is to remain a free variable, independent of the scale $k$, it follows from Eq.~\eqref{eq:phidef} that
\begin{equation}
\label{eq:freephi}
\partial_k\phi_x=-\partial_k\frac{\delta W[\mathcal{J},\mathcal{R}^{(k)}]}{\delta \mathcal{J}_x}\overset{!}{=}0,
\end{equation}
and $\mathcal{J}_x\equiv\mathcal{J}_x^{(k)}[\phi]$ must therefore be a function of $k$. Varying Eq.~\eqref{eq:ave1PI} with respect to the scale $k$ then gives
\begin{align}
\label{eq:Wetterichstart}
	\partial_k \Gamma^{\rm 1PI}_{\rm av}[\phi,\mathcal{R}^{(k)}]=&\,\partial_k W[\mathcal{J}^{(k)},\mathcal{R}^{(k)}]+\phi_x\partial_k\mathcal{J}_x^{(k)}\nonumber\\&+\frac{1}{2}\phi_x\partial_k\mathcal{R}^{(k)}_{xy}\phi_y,
\end{align}
and the derivative of the Schwinger functional is
\begin{align}
\label{eq:Wvary}
\partial_k W[\mathcal{J}^{(k)},\mathcal{R}^{(k)}]=&-\phi_x\partial_k\mathcal{J}_x^{(k)}\nonumber\\&-\frac{1}{2}\left(\hbar\Delta^{\!(k)}_{xy}+\phi_x\phi_y\right)\partial_k\mathcal{R}^{(k)}_{xy},
\end{align}
where we have defined the connected two-point function
\begin{equation}
\Delta^{(k)}_{xy}=-\frac{\delta^2 W[\mathcal{J}^{(k)},\mathcal{R}^{(k)}]}{\delta \mathcal{J}^{(k)}_x\delta \mathcal{J}^{(k)}_y}.
\end{equation}
Substituting Eq.~\eqref{eq:Wvary} back into Eq.~\eqref{eq:Wetterichstart}, we obtain
\begin{equation}
\label{eq:Wetterich}
\partial_k\Gamma^{\rm 1PI}_{\rm av}[\phi,\mathcal{R}^{(k)}]=-\frac{\hbar}{2}\,\mathrm{Tr}\left(\Delta^{\!(k)} \ast \partial_k \mathcal{R}^{(k)}\right),
\end{equation}
where the asterisk indicates a spacetime convolution, i.e.~\smash{$\Delta^{\!(k)}\ast\mathcal{R}^{(k)}\equiv \Delta^{\!(k)}_{xy}\mathcal{R}^{(k)}_{yz}$}. Equation~\eqref{eq:Wetterich} is the well-known flow equation of the functional RG.

\subsection{2PI approach}

All of the information about the dynamics of an interacting system is encoded in the infinite set of its $n$-point functions, and the coupled system of equations that these functions satisfy can be derived from the $n$PI effective action. In the case of the RG flow, we are interested in knowing how this set of $n$-point functions changes with scale. Since the flow equation of the functional RG is concerned with the change in the two-point function with scale, it seems reasonable therefore that the starting point should be the 2PI effective action.

Before proceeding, it is important to consider the convexity of the 2PI and average 1PI effective actions. It is well known that the $n$PI effective actions are convex with respect to the variables that are convex conjugate to the sources. In the case of the 2PI effective action, the convex-conjugate variables are $\phi'_x\equiv\phi_x$ and $\Delta_{xy}'\equiv\hbar\Delta_{xy}+\phi_x\phi_y$, see Ref.~\cite{Millington:2019nkw}. Specifically, we have
\begin{equation}
\label{eq:Hess}
-{\rm Hess}(\Gamma^{\rm 2PI})(\phi',\Delta')\cdot {\rm Hess}(W)(\mathcal{J}',\mathcal{K}')=\mathbb{I},
\end{equation}
where ${\rm Hess}(\Gamma^{\rm 2PI})(\phi',\Delta^{\prime})$ is the functional Hessian matrix of $\Gamma^{\rm 2PI}[\phi,\Delta]$ with respect to $\phi'$ and $\Delta'$, and ${\rm Hess}(W)(\mathcal{J}',\mathcal{K}')$ is the functional Hessian matrix of $W[\mathcal{J},\mathcal{K}]$ with respect to $\mathcal{J}'\equiv\mathcal{J}$ and $\mathcal{K}^{\prime}\equiv\mathcal{K}/2$. From Eq.~\eqref{eq:vardefs}, we see that ${\rm Hess}(W)(\mathcal{J}',\mathcal{K}')$ is the negative of a covariance matrix and therefore negative semi-definite. Excepting the singular case, it follows from Eq.~\eqref{eq:Hess} that ${\rm Hess}(\Gamma^{\rm 2PI})(\phi',\Delta^{\prime})$ is positive definite, implying that $\Gamma^{\rm 2PI}[\phi,\Delta]$ is convex with respect to $\phi'_x$ and $\Delta_{xy}'$. However, for a given \smash{$\Delta_{xy}\neq\Delta'_{xy}$}, the 2PI effective action need not be convex in the $\phi$ direction. This is shown explicitly in the case of a zero-dimensional quantum field theory with spontaneous symmetry breaking in Ref.~\cite{Millington:2019nkw} (see Fig.~3(a) therein). Hence, as is true of the average 1PI effective action, the 2PI effective action for a given $\Delta_{xy}$ is not, in general, convex in the $\phi$ direction, as required for it to yield consistent RG evolution.

Returning to the 2PI effective action, its variation with respect to the scale $k$ is given by
\begin{equation}
\partial_k\Gamma^{\rm 2PI}[\phi,\Delta]=\frac{\delta \Gamma^{\rm 2PI}[\phi,\Delta]}{\delta \phi_x}\,\partial_k\phi_x+\frac{\delta \Gamma^{\rm 2PI}[\phi,\Delta]}{\delta \Delta_{xy}}\,\partial_k\Delta_{xy}.
\end{equation}
Again imposing that
\begin{equation}
\label{eq:partialphi}
\partial_k\phi_x=-\partial_k\frac{\delta W[\mathcal{J},\mathcal{K}]}{\delta \mathcal{J}_x}=0,
\end{equation}
and making use of Eq.~\eqref{eq:JplKdef}, we have that
\begin{equation}
\partial_k\Gamma^{\rm 2PI}[\phi,\Delta]=\frac{\hbar}{2}\,\mathcal{K}_{xy}[\phi,\Delta]\partial_k\Delta_{xy}.
\end{equation}
Choosing $\mathcal{K}_{xy}[\phi,\Delta]\equiv\mathcal{K}_{xy}^{(k)}[\phi,\Delta]=\mathcal{R}^{(k)}_{xy}$ to be the regulator, Eq.~\eqref{eq:partialphi} fixes $\mathcal{J}_x\equiv\mathcal{J}_{x}^{(k)}[\phi,\Delta]$, and we obtain
\begin{equation}
\partial_k\Gamma^{\rm 2PI}[\phi,\Delta]=\frac{\hbar}{2}\,\mathrm{Tr}\left(\mathcal{R}^{(k)}\ast\partial_k\Delta\right).
\end{equation}

We emphasize that the above restriction of the sources $\mathcal{J}_x$ and $\mathcal{K}_{xy}$ fixes the two-point function \smash{$\Delta_{xy}\equiv\Delta_{xy}^{(k)}$} to be a functional of $\phi$ (see Ref.~\cite{Millington:2021ftp}), and the flow remains closed as in the average 1PI approach. This follows directly from the convexity of the 2PI effective action, which allows us to write
\begin{subequations}
\label{eq:fromconvex}
\begin{align}
&\frac{\delta^2\Gamma^{\rm 2PI}}{\delta \phi'_x\delta \phi'_z}\,\frac{\delta^2W}{\delta \mathcal{J}^{\prime(k)}_z\delta \mathcal{J}^{\prime(k)}_y}+\frac{\delta^2 \Gamma^{\rm 2PI}}{\delta \phi'_x\delta \Delta^{\prime(k)}_{zw}}\,\frac{\delta^2W}{\delta \mathcal{K}^{\prime(k)}_{zw}\delta \mathcal{J}^{\prime(k)}_y}=-\delta^{(d)}_{xy},\\
&\frac{\delta^2\Gamma^{\rm 2PI}}{\delta \phi_x'\delta \phi'_u}\frac{\delta^2W}{\delta \mathcal{J}^{\prime(k)}_u\delta\mathcal{K}^{\prime(k)}_{yz}}+\frac{\delta^2\Gamma^{\rm 2PI}}{\delta\phi_x'\delta\Delta^{\prime(k)}_{uv}}\frac{\delta^2W}{\delta \mathcal{K}^{\prime(k)}_{uv}\delta\mathcal{K}^{\prime(k)}_{yz}}=0,\\
&\frac{\delta^2\Gamma^{\rm 2PI}}{\delta \Delta^{\prime(k)}_{xy}\delta \phi'_u}\frac{\delta^2W}{\delta \mathcal{J}^{\prime(k)}_u\delta\mathcal{J}^{\prime(k)}_{z}}+\frac{\delta^2\Gamma^{\rm 2PI}}{\delta\Delta^{\prime(k)}_{xy}\delta\Delta^{\prime(k)}_{uv}}\frac{\delta^2W}{\delta \mathcal{K}^{\prime(k)}_{uv}\delta\mathcal{J}^{\prime(k)}_{z}}=0,\end{align}
\begin{align}
&\frac{\delta^2\Gamma^{\rm 2PI}}{\delta \Delta^{\prime(k)}_{xy}\delta\Delta^{\prime(k)}_{uv}}\frac{\delta^2W}{\delta\mathcal{K}^{\prime(k)}_{uv}\delta\mathcal{K}^{\prime(k)}_{zw}}+\frac{\delta^2\Gamma^{\rm 2PI}}{\delta \Delta^{\prime(k)}_{xy}\delta\phi'_u}\frac{\delta^2W}{\delta\mathcal{J}^{\prime(k)}_u\delta\mathcal{K}^{\prime(k)}_{zw}}\nonumber\\&\qquad=-\frac{1}{2}\left(\delta^{(d)}_{xz}\delta^{(d)}_{yw}+\delta^{(d)}_{xw}\delta^{(d)}_{yz}\right),
\end{align}
\end{subequations}
in $d$ spacetime dimensions, where $\Gamma^{\rm 2PI}\equiv \Gamma^{\rm 2PI}[\phi,\Delta^{\!(k)}]$ and $W\equiv W[\mathcal{J}^{(k)},\mathcal{K}^{(k)}]$.
Using
\begin{subequations}
\begin{align}
    \frac{\delta}{\delta \phi'_x}&=\frac{\delta}{\delta \phi_x}-\frac{2}{\hbar}\phi_y\frac{\delta}{\delta \Delta^{(k)}_{yx}},\\
    \frac{\delta}{\delta \Delta^{\prime(k)}_{xy}}&=\frac{1}{\hbar}\frac{\delta}{\delta \Delta^{(k)}_{xy}},
\end{align}
\end{subequations}
as well as the other identities collected in the third to sixth rows of Tab.~\ref{tab:comp}, Eq.~\eqref{eq:fromconvex} can be written (see Ref.~\cite{Millington:2021ftp})
\begin{widetext}
\begin{subequations}
\label{eq:fromconvex2}
\begin{align}
    \label{eq:givesDeltainv}
    &\left\{\frac{\delta^2 \Gamma^{\rm 2PI}}{\delta \phi_x\delta \phi_z}-\mathcal{K}_{xz}^{(k)}-\frac{4\phi_w}{\hbar}\left[\frac{\delta^2\Gamma^{\rm 2PI}}{\delta\phi_{(x}\delta\Delta_{z)w}}-\frac{\phi_u}{\hbar}\frac{\delta^2\Gamma^{\rm 2PI}}{\delta \Delta_{wx}\delta\Delta_{uz}}\right]\!\right\}\Delta_{zy}
    -\frac{2}{\hbar}\left[\frac{\delta^2 \Gamma^{\rm 2PI}}{\delta \phi_x\delta\Delta_{zw}}-\frac{2\phi_u}{\hbar}\frac{\delta^2 \Gamma^{\rm 2PI}}{\delta\Delta_{ux}\delta\Delta_{zw}}\right]\!\frac{\delta^2W}{\delta\mathcal{J}_y\delta \mathcal{K}_{zw}}=\delta^{(d)}_{xy},\\
    &\left\{\frac{\delta^2 \Gamma^{\rm 2PI}}{\delta \phi_x\delta \phi_z}-\mathcal{K}_{xz}^{(k)}-\frac{4\phi_w}{\hbar}\left[\frac{\delta^2\Gamma^{\rm 2PI}}{\delta\phi_{(x}\delta\Delta_{z)w}}-\frac{\phi_u}{\hbar}\frac{\delta^2\Gamma^{\rm 2PI}}{\delta \Delta_{wx}\delta\Delta_{uz}}\right]\!\right\}\frac{\delta^2W}{\delta\mathcal{J}_z\delta \mathcal{K}_{vy}}
    +\frac{2}{\hbar}\left[\frac{\delta^2 \Gamma^{\rm 2PI}}{\delta \phi_x\delta\Delta_{zw}}-\frac{2\phi_u}{\hbar}\frac{\delta^2 \Gamma^{\rm 2PI}}{\delta\Delta_{ux}\delta\Delta_{zw}}\right]\!\frac{\delta^2W}{\delta \mathcal{K}_{vy}\delta \mathcal{K}_{zw}}=0,\\
    \label{eq:conv3}
    &\left[\frac{\delta^2 \Gamma^{\rm 2PI}}{\delta \phi_w\delta\Delta_{xy}}-\frac{2\phi_u}{\hbar}\frac{\delta^2 \Gamma^{\rm 2PI}}{\delta\Delta_{uw}\delta\Delta_{xy}}\right]\Delta_{wz}
    -\frac{2}{\hbar}\frac{\delta^2\Gamma^{\rm 2PI}}{\delta \Delta_{wu}\delta \Delta_{xy}}\frac{\delta^2W}{\delta\mathcal{J}_z\delta \mathcal{K}_{wu}}=0,\\
    &\,\frac{2}{\hbar}\left\{\left[\frac{\delta^2\Gamma^{\rm 2PI}}{\delta \phi_u\delta \Delta_{xy}}-\frac{2\phi_v}{\hbar}\frac{\delta^2 \Gamma^{\rm 2PI}}{\delta \Delta_{vu}\delta\Delta_{xy}}\right]\frac{\delta^2W}{\delta\mathcal{J}_u\delta \mathcal{K}_{zw}}
    +\frac{2}{\hbar}\frac{\delta^2\Gamma^{\rm 2PI}}{\delta \Delta_{xy}\delta \Delta_{uv}}\frac{\delta^2W}{\delta \mathcal{K}_{uv}\delta \mathcal{K}_{zw}}\right\}=-\frac{1}{2}\left(\delta^{(d)}_{xz}\delta^{(d)}_{yw}+\delta^{(d)}_{xw}\delta^{(d)}_{yz}\right),
\end{align}
\end{subequations}
\end{widetext}
 as appeared (in condensed notation) in footnote 11 of Ref.~\cite{Cornwall:1974vz}.  Herein, we have omitted the superscript ``$(k)$'' on $\Delta$, $\mathcal{J}$ and $\mathcal{K}$, and used the shorthand notation \smash{$A_{(x}B_{z)w}\equiv\frac{1}{2}\left(A_{x}B_{zw}+A_{z}B_{xw}\right)$} for symmetrization in the coordinates.
 
By solving Eq.~\eqref{eq:fromconvex2}, we can show (see Ref.~\cite{Millington:2021ftp}) that the inverse two-point function is given by
\begin{align}
    \label{eq:Deltaexplicit}
    \Delta^{(k),-1}_{xy}&=\frac{\delta^2 \Gamma^{\rm 2PI}}{\delta \phi_x\delta \phi_y}-\mathcal{K}^{(k)}_{xy}\nonumber\\&\qquad-\frac{\delta^2\Gamma^{\rm 2PI}}{\delta \phi_x\delta \Delta^{(k)}_{zw}}\left(\frac{\delta^2\Gamma^{\rm 2PI}}{\partial \Delta^{(k)}_{zw}\delta \Delta^{(k)}_{uv}}\right)^{-1}\frac{\delta^2\Gamma^{\rm 2PI}}{\delta \Delta^{(k)}_{uv}\delta \phi_y}\nonumber\\&=S^{(2)}_{xy}[\phi]-\mathcal{K}^{(k)}_{xy}+\mathcal{O}(\hbar).
\end{align}
where
\begin{equation}
\label{eq:Sn}
S^{(2)}_{xy}[\phi]\equiv\left.\frac{\delta^2 S[\Phi]}{\delta \Phi_x\delta \Phi_y}\right|_{\Phi=\phi}.
\end{equation}

\begin{table*}[t!]
\begin{center}
\everymath{\displaystyle}
\footnotesize
\begin{tabular}{| c | c |}\hline
Average 1PI & 2PI \\ \hline \hline
$\Gamma^{\rm 1PI}_{\rm av}[\phi,\mathcal{R}^{(k)}]=W[\mathcal{J}^{(k)}[\phi],\mathcal{R}^{(k)}]+\mathcal{J}^{(k)}_x[\phi]\phi_x+\tfrac{1}{2}\mathcal{R}_{xy}^{(k)}\phi_x\phi_y$ & $\begin{aligned}\Gamma^{\rm 2PI}[\phi,\Delta^{\!(k)}]&=W[\mathcal{J}^{(k)}[\phi,\Delta^{\!(k)}],\mathcal{K}^{(k)}[\phi,\Delta^{\!(k)}]]\\[-0.3em]
&\phantom{=}+\mathcal{J}_x[\phi,\Delta^{\!(k)}]\phi_x+\tfrac{1}{2}\mathcal{K}_{xy}^{(k)}[\phi,\Delta^{\!(k)}]\left(\phi_x\phi_y+\hbar \Delta_{xy}^{\!(k)}\right)\end{aligned}$\\ \hline
$\phi_x=-\frac{\delta W[\mathcal{J}^{(k)}[\phi],\mathcal{R}^{(k)}]}{\delta \mathcal{J}^{(k)}_x[\phi]}$ & $\phi_x=-\frac{\delta W[\mathcal{J}^{(k)}[\phi,\Delta^{\!(k)}],\mathcal{K}^{(k)}[\phi,\Delta^{\!(k)}]]}{\delta \mathcal{J}^{(k)}_x[\phi,\Delta^{\!(k)}]}$\\ \hline
$\begin{aligned}\hbar \Delta^{\!(k)}_{xy}&=-2\frac{\delta W[\mathcal{J}^{(k)}[\phi],\mathcal{R}^{(k)}]}{\delta \mathcal{R}_{xy}^{(k)}}-\phi_x\phi_y\\[-0.2em]&=-\hbar\frac{\delta^2 W[\mathcal{J}^{(k)}[\phi],\mathcal{R}^{(k)}]}{\delta \mathcal{J}^{(k)}_x[\phi]\delta \mathcal{J}^{(k)}_y[\phi]}\end{aligned}$ & $\begin{aligned}\hbar \Delta^{\!(k)}_{xy}&=-2\frac{\delta W[\mathcal{J}^{(k)}[\phi,\Delta^{\!(k)}],\mathcal{K}^{(k)}[\phi,\Delta^{\!(k)}]]}{\delta \mathcal{K}_{xy}^{(k)}[\phi,\Delta^{\!(k)}]}-\phi_x\phi_y\\[-0.2em]&=-\hbar\frac{\delta^2 W[\mathcal{J}^{(k)}[\phi,\Delta^{\!(k)}],\mathcal{K}^{(k)}[\phi,\Delta^{\!(k)}]]}{\delta \mathcal{J}^{(k)}_x[\phi,\Delta^{\!(k)}]\delta \mathcal{J}^{(k)}_y[\phi,\Delta^{\!(k)}]}\end{aligned}$\\ \hline
$\frac{\delta \Gamma^{\rm 1PI}_{\rm av}[\phi,\mathcal{R}^{(k)}]}{\delta \phi_x}=\mathcal{J}^{(k)}_x[\phi]+\mathcal{R}^{(k)}_{xy}\phi_y$ & $\frac{\delta \Gamma^{\rm 2PI}[\phi,\Delta^{\!(k)}]}{\delta \phi_x}=\mathcal{J}_x^{(k)}[\phi,\Delta^{\!(k)}]+\mathcal{K}_{xy}^{(k)}[\phi,\Delta^{\!(k)}]\phi_y$\\ \hline $\frac{\delta \Gamma^{\rm 1PI}_{\rm av}[\phi,\mathcal{R}^{(k)}]}{\delta \mathcal{R}_{xy}^{(k)}}=-\tfrac{\hbar}{2} \Delta^{\!(k)}_{xy}$ & $\frac{\delta \Gamma^{\rm 2PI}[\phi,\Delta^{\!(k)}]}{\delta \Delta^{\!(k)}_{xy}}=+\tfrac{\hbar}{2}\mathcal{K}_{xy}^{(k)}[\phi,\Delta^{\!(k)}]$\\ \hline
$\begin{aligned}
\Delta^{\!(k),-1}_{xy}&=\frac{\delta^2\Gamma^{\rm 1PI}[\phi,\mathcal{R}^{\!(k)}]}{\delta\phi_x\delta\phi_y}\\&=\frac{\delta^2\Gamma^{\rm 1PI}_{\rm av}[\phi,\mathcal{R}^{\!(k)}]}{\delta\phi_x\delta\phi_y}-\mathcal{R}^{(k)}_{xy}\\&=S^{(2)}_{xy}[\phi]-\mathcal{R}_{xy}^{(k)}+\mathcal{O}(\hbar)\end{aligned}$ & $\begin{aligned}\Delta^{(k),-1}_{xy}&=\frac{\delta^2 \Gamma^{\rm 2PI}[\phi,\Delta^{(k)}]}{\delta \phi_x\delta \phi_y}-\mathcal{K}^{(k)}_{xy}[\phi,\Delta^{(k)}]\\&-\frac{\delta^2\Gamma^{\rm 2PI}[\phi,\Delta^{(k)}]}{\delta \phi_x\delta \Delta_{zw}^{(k)}}\left(\frac{\delta^2\Gamma^{\rm 2PI}[\phi,\Delta^{(k)}]}{\partial \Delta^{(k)}_{zw}\delta \Delta^{(k)}_{uv}}\right)^{-1}\frac{\delta^2\Gamma^{\rm 2PI}[\phi,\Delta^{(k)}]}{\delta \Delta^{(k)}_{uv}\delta \phi_y}\\&=S_{xy}^{(2)}[\phi]-\mathcal{K}_{xy}^{(k)}[\phi,\Delta^{\!(k)}]+\mathcal{O}(\hbar)\end{aligned}$\\ \hline
$\partial_k\Gamma^{\rm 1PI}_{\rm av}[\phi,\mathcal{R}^{(k)}]=\frac{\delta \Gamma^{\rm 1PI}_{\rm av}[\phi,\mathcal{R}^{(k)}]}{\delta \phi_x}\,\partial_k\phi_x+\frac{\delta \Gamma^{\rm 1PI}_{\rm av}[\phi,\mathcal{R}^{(k)}]}{\delta \mathcal{R}_{xy}^{(k)}}\,\partial_k\mathcal{R}_{xy}^{(k)}$ & $\partial_k\Gamma^{\rm 2PI}[\phi,\Delta^{\!(k)}]=\frac{\delta \Gamma^{\rm 2PI}[\phi,\Delta^{\!(k)}]}{\delta \phi_x}\,\partial_k\phi_x+\frac{\delta \Gamma^{\rm 2PI}[\phi,\Delta^{\!(k)}]}{\delta \Delta^{\!(k)}_{xy}}\,\partial_k\Delta^{\!(k)}_{xy}$ \\ \hline
\end{tabular}
\end{center}
\caption{\label{tab:comp} Comparison of the average 1PI and 2PI effective actions, and their variations. Notice that \smash{$\Delta^{\!(k),-1}$} is obtained from the functional variation of the shifted average 1PI effective action \smash{$\Gamma^{\rm 1PI}[\phi,\mathcal{R}^{(k)}]\equiv\Gamma^{\rm 1PI}_{\rm av}[\phi,\mathcal{R}^{(k)}]-\frac{1}{2}\phi_x\mathcal{R}_{xy}^{(k)}\phi_y$} in the standard approach. The functional dependencies of the sources have been included explicitly for clarity. We draw attention to the interchange of the roles played by \smash{$\mathcal{R}^{(k)}_{xy}\equiv \mathcal{K}^{(k)}_{xy}[\phi,\Delta]$} and \smash{$\Delta_{xy}^{\!(k)}$} in the sixth and eighth rows due to the additional Legendre transform between the left and right columns.}
\end{table*}

The above approach differs from that of Ref.~\cite{Lavrov:2012xz}, which was motivated by problems of on-shell gauge dependence in the average effective action. In particular, following Ref.~\cite{Lavrov:2012xz} would amount here to the introduction of a source conjugate to the composite operator \smash{$\phi_x\mathcal{R}_{xy}^{(k)}\phi_y$}.  We remark, however, that the method of external sources~\cite{Garbrecht:2015cla}, upon which our approach is based, was introduced as a way of ensuring that symmetry properties can be preserved in truncations of the 2PI effective action. We anticipate that our alternative derivation of the flow equation can readily be extended to include the additional contributions to the sources needed to preserve symmetry properties and we leave it for further work to show whether this methodology can also be used to alleviate problems of gauge dependence.

\section{Discussion}

The average-1PI and 2PI procedures that we have described lead to two distinct flow equations:
\begin{subequations}
\label{eq:flows}
\begin{align}
\label{eq:1PIflow}
\partial_k\Gamma^{\rm 1PI}_{\rm av}[\phi,\mathcal{R}^{(k)}] = -\frac{\hbar}{2}\,{\rm STr}\left(\Delta^{\!(k)}\ast \partial_k\mathcal{R}^{(k)}\right),\\
\label{eq:2PIflow}
\partial_k\Gamma^{\rm 2PI}[\phi,\Delta^{\!(k)}] = +\frac{\hbar}{2}\,{\rm STr}\left(\mathcal{R}^{(k)}\ast \partial_k\Delta^{\!(k)}\right),
\end{align}
\end{subequations}
wherein we have promoted the trace to a supertrace over the spacetime indices and any additional internal indices for generality. In Eq.~\eqref{eq:1PIflow}, the flow of the effective action depends \emph{directly} on the scale dependence of the regulator. In Eq.~\eqref{eq:2PIflow}, the flow of the effective action instead depends only \emph{indirectly} on the scale dependence of the regulator, through the scale dependence of the two-point function. In other words, the introduction of the regulator \emph{always} causes a flow of the average 1PI effective action, but the 2PI effective action flows only if the two-point function responds to the regulator.

In order to go from the average 1PI effective action to the 2PI effective action, we must perform an additional Legendre transform, adding to the former a term \smash{$\hbar\mathcal{R}^{(k)}_{xy}\Delta^{\!(k)}_{yx}$/2}. The variation of this term with the scale $k$ accounts for the difference between the right-hand sides of Eqs.~\eqref{eq:1PIflow} and~\eqref{eq:2PIflow}. A comparison of the two procedures is given in Tab.~\ref{tab:comp}. The two results coincide if
\begin{equation}
\label{eq:diff}
\partial_k{\rm STr}\left(\Delta^{\!(k)} \ast\mathcal{R}^{(k)}\right)=0,
\end{equation}
and this is not, in general, the case. 

Returning to Eq.~\eqref{eq:diff}, and making use of Eqs.~\eqref{eq:Deltaexplicit} and~\eqref{eq:Sn}, we can write
\begin{align}
\Delta^{(k)}_{xy}\partial_k\mathcal{R}^{(k)}_{yx}=-\Delta^{\!(k)}_{xy}\partial_{k}\Delta^{\!(k),-1}_{yx}+\mathcal{O}(\hbar).
\end{align}
Since $\partial_{k}\Delta_{yx}^{\!(k),-1}=-\Delta_{yz}^{\!(k),-1}\left(\partial_k\Delta^{\!(k)}_{zw}\right)\Delta^{\!(k),-1}_{wx}$, we find
\begin{equation}
\label{eq:corrections}
\partial_k\left(\Delta^{\!(k)}_{xy}\mathcal{R}^{(k)}_{yx}\right)=S^{(2)}_{xy}[\phi]\partial_k\Delta^{\!(k)}_{yx}+\mathcal{O}(\hbar),
\end{equation}
which is, in general, nonzero, such that there is a material difference between the flow equations in Eq.~\eqref{eq:flows}. The first term on the right-hand side of Eq.~\eqref{eq:corrections} can be seen as a correction to the Wetterich-Morris-Ellwanger equation.

Lastly, we consider the boundary conditions on the 2PI and average 1PI effective actions. In the limit $k\to 0$, the regulator vanishes ($\mathcal{R}^{(k)}\to 0$). In this case, both the 2PI and average 1PI effective actions coincide with the 1PI effective action $\Gamma^{\rm 1PI}[\phi]=W[\mathcal{J}]+\mathcal{J}_x\phi_x$. (The 2PI effective action for a vanishing two-point source is precisely the 1PI effective action.) Instead, for $k\to\infty$, and if the regulator diverges (i.e., $\mathcal{R}^{(k)}\to -\infty$) in the same limit, we do not integrate in any fluctuations, and both the 2PI and average 1PI effective actions coincide with the classical action $S$ (see Ref.~\cite{Berges:2000ew}). Thus, both approaches share the same boundary conditions, and the only difference is in the form of the corresponding flow equations [Eq.~\eqref{eq:flows}].

\section{Concluding remarks}

In summary, we have derived an alternative flow equation for the functional RG evolution, which differs from the Wetterich-Morris-Ellwanger equation. While the former is derived from the average 1PI effective action, we have instead employed a self-consistent procedure based on the 2PI effective action. An extended discussion of the differences in the resulting RG evolution for the $\lambda\phi^4$ theory is presented in a follow-up work~\cite{Alexander:2019quf}.

%%%%%%%%%%%%%%%%%%%%%%%%%%%%%%%%%%%%%%%%%%%%%%%%%%%%%%%%%%%%%%%%%%%%%%

\begin{acknowledgments}
This work is based in part on the masters dissertations of EA and JN, supervised by PM in the School of Physics and Astronomy at the University of Nottingham. PM would like to thank Bj\"{o}rn Garbrecht for many enjoyable discussions and earlier collaboration in this area. The authors would like to thank Jean Alexandre and Tim Morris for comments on the manuscript, and the participants of the International Seminar on Asymptotic Safety for their engagement with this work. The Authors also thank Dario Benedetti, Kevin Falls, Jan Pawlowski and Adam Rancon for discussions at the 10th International Conference on Exact Renormalizatiomn Group 2020 (ERG2020), hosted by the Yukawa Institute for Theoretical Physics, Japan, which led to an important correction to the expression for the inverse two-point function, appearing here and in an erratum, submitted in respect of the published article. This work was supported by a Nottingham Research Fellowship from the University of Nottingham; the Leverhulme Trust [grant number RL-2016-028]; and the Science and Technology Facilities Council [grant number ST/P000703/1].
\end{acknowledgments}

\end{document}